\documentclass[12pt,a4paper,final]{iopart}

\usepackage{caption}
\usepackage{float}
\usepackage{iopams}  
\usepackage{graphicx}
\usepackage[breaklinks=true,colorlinks=true,linkcolor=blue,urlcolor=blue,citecolor=blue]{hyperref}
\hypersetup{colorlinks=true, urlcolor=blue, citecolor=blue}
\begin{document}

\title{Shifting of Fermi Level and Realization of Topological Insulating Phase in the Oxyfluoride BaBiO$_2$F}

\author{Bramhachari Khamari and B. R. K Nanda$^1$}
\address{$^1$ Condensed Matter Theory and Computational Lab, Department of Physics, Indian Institute of Technology Madras, Chennai 600036, India}


\eads{\mailto{nandab@iitm.ac.in}} 

\begin{abstract}
The disadvantage of BaBiO$_3$ of not being a topological insulator despite having symmetry protected Dirac state is overcome by shifting the Fermi level (E$_F$) via fluorination. The DFT calculations reveal that the fluorination neither affects the spin-orbit coupling nor the parity of the states, but it acts as a perfect electron donor to shift the E$_F$. We find that 33 \% fluorination is sufficient to shift the E$_F$ by $\sim$ 2 eV so that the invariant Dirac state lies on it to make BaBiO$_2$F a topological insulator. The fluorinated cubic compound can be experimentally synthesized as the phonon studies predict dynamical stability above $\sim$ 500 K. Furthermore, the Dirac states are found to be invariant against the low-temperature phase lattice distortion which makes the structure monoclinic. The results carry practical significance as they open up the possibility of converting the family of superconducting oxides, ABiO$_3$ (A = Na, K, Cs, Ba, Sr, Ca), to real topological insulator through appropriate fluorination.

\end{abstract}

\section{Introduction}
Band topology, one of the highly active research area in materials science and physical chemistry for the last one decade, has given rise to the novel exotic concept of topological insulator (TI) \cite{Hsieh,Liang,Hasan,Wray,Pal,Sato,Kubler,Moore,Hughes}. The TI compounds with an insulating interior and conducting exterior \cite{Hsieh,Wray}, and determined through Z$_2$ invariance\cite{Fu}, are  promising for next generation spin based device applications such as spintronics and quantum information and quantum computing \cite{Shuichi,LiangFu}. The spin-orbit coupling (SOC) driven complex band structure of these compounds, is a great recipe to realize high thermoelectric efficiency\cite{David}.  The conducting exterior of the TIs exhibit topologically invariant Dirac cones (TIDC), where linearly dispersing valence band (VB) and conduction band (CB) touch each other at the Fermi level (E$_F$) and are protected by time reversal symmetry. 

While the widely investigated topological insulators, namely selenides, tellurides\cite{Shou,Cheng,Park} and Heusler\cite{Kubler,Jun} alloys,  have demonstrated interesting physical concepts, the narrow band gap ($\sim$ 0.3eV) hinders their application prospects\cite{Clau,Binghai} and hence search for wide band gap TIs is highly desirable. In addition, from the research point of view, wide-band gap 3D systems exhibiting high resistance are useful to experimentally reveal the TI properties, as the surface and bulk states can be easily distinguished\cite{Jin}. In this context, covalent bonding driven large band gap oxides should be of great interest. However, due to negligible contribution towards SOC by the oxygen states, many of the oxides fail to exhibit the TI behavior.

There are very few oxides, such as Bi based double perovskites {\it A}$_2$BiXO$_6$, where {\it A} is a group-II element and X is either Br and I and single perovskites ABiO$_3$ with A being a group-I and II element, B being a group-V and VI heavy element which exhibit topologically protected surface Dirac states in their equilibrium structure \cite{Pi,Clau,Thomale,Nanda}. Since  the compounds Ba$_{1-x}$K$_x$BiO$_3$ are superconductors (T$_C$ $\sim$ 30K\cite{Cava}), it is suggested that both the superconducting behavior and the topologically invariant surface states\cite{Thomale}can together be exploited to realize Majorona Fermion in these matrials\cite{Kanecl}. The perovskites also carry significance, as their layered geometry facilitates to construct heterostructures so that the competition between the topological phase and various orders can be examined to explore novel interfacial states\cite{Hwan,Linder}. However, the major disadvantage of ABiO$_3$ is that the bulk band gap ($\sim$ 0.7 eV) within which the surface TIDC appears, lies away from the Fermi level (E$_F$). The separation between TIDC and E$_F$ is in the range 2 to 5 eV\cite{Nanda}. Hence, for all practical purposes, these compounds are not TI,  unless the E$_F$ is shifted to the TIDC. The shifting of E$_F$ is usually achieved  through certain mechanism like thermal excitation, electrical gating or chemical doping. While thermal excitation can hardly shift the E$_F$ by few meV, the electrical gating can push it to a maximum of 1 eV\cite{Qiran}.  Therefore, chemical substitution remains the only viable option though it can distort the lattice to lower the symmetry and thereby to affect the parity of the states which determine Z$_2$ invariance. In ABiO$_3$, while O and A have fixed charge states, 2$^-$, 1$^+$/2$^+$ respectively, Bi is multivalent and show ambiguous (flexible) charge states\cite{Nanda}.  Hence, the chemical doping is preferable either at the A site or at the anion site so that the charge neutrality is maintained.  

In this work, with the objective of realizing the TIDC at E$_F$ through the substitution, we have examined the electronic structure of BaBiO$_3$ as a prototype. 
In order to achieve the electron doping, we replace O by F. Fluorine is a preferable substituent as studies on other oxides suggest that it maintains the thermodynamical stability while replacing the oxygen. For example, F is substituted for O to achieve electron doping in the iron based superconductor, LaO$_{0.5}$F$_{0.5}$BiS$_{2}$ (x= 0.5)\cite{Usui,Lee}, LaO$_{1-x}$F$_{x}$FeAs (x= 0.05 $-$ 0.12)\cite{Takumi}. Similarly, experimental reports suggest the formation of La$_{1-x}$Sr$_{x}$FeO$_{3-x}$F$_{x}$ (x=1, 0.8, 0.5, 0.2)\cite{Clemens}. Furthermore, it has been found that perovskites BaFeO$_2$F \cite{Heap,Frank} and SrFeO$_2$F \cite{Thompson,Shim} can be experimentally synthesized. In these perovskites, one out of three Oxygens are replaced with Fluorine.

\begin{figure}
\begin{center}
\hspace{0.1cm}
\includegraphics[angle=-0.0,origin=c,height=6.5cm,width=8.0cm]{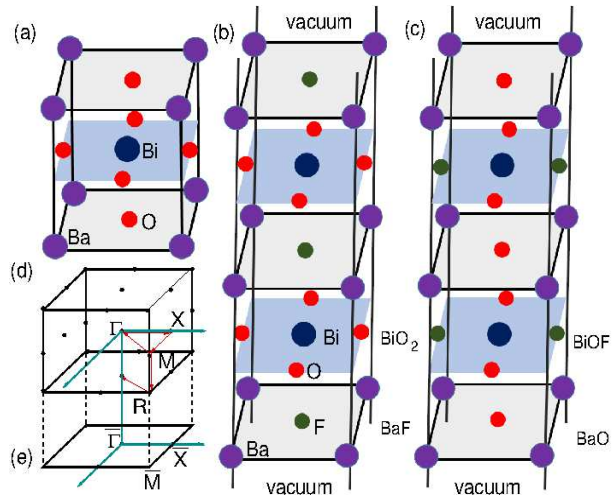}
\caption{(a) The cubic perovskite structure of BaBiO$_3$. (b - c) Representative structures of the [001] BaBiO$_2$F slabs with F being placed at different face centered positions. The real slabs, on which the calculations were carried out, consist of 15 units of BaBiO$_2$F with 15 {\AA} thick vacuum separating two consecutive slabs. (d) and (e) depict the bulk and surface Brillouin zones (BZ) respectively. For these systems, the time reversal invariant momenta (TRIM) are point R and $\Gamma$ which are mapped to $\overline{M}$ and $\overline{\Gamma}$ respectively in the surface BZ.} 
\label{fig:1}
\end{center}
\end{figure}

The compound BaBiO$_3$ undergoes a structural transition from cubic to rhombohedral to monoclinic phase with decrease in the temperature\cite{Kennedy} and the earlier studies suggest that the TIDC states are protected against these structural transitions\cite{Clau}. Therefore, as the high-symmetry configurations are computationally less expensive, in this work we deal with the high temperature(750 K-800 K) cubic phase of BaBiO$_3$\cite{Sle,Cox}. The crystal structure, the slab geometry and their corresponding Brillouin zones (BZ) are shown in Fig. 1. 

To examine the band topology of the pure and fluorinated compounds, density functional calculations are carried out using the full potential linearized plane wave (FP-LAPW) formalism \cite{Hamann} as implemented in WIEN2k simulation package\cite{Aug}.
Augmented plane waves in the interstitial and localised orbitals within the muffin-tin sphere are used to construct the basis sets. To account for exchange and correlation effect generalized gradient approximation (GGA) \cite{Perdew} along with the modified Becke-Johnson (mBJ) correction\cite{Tran} is considered. For the  the BZ integration, $10\times{10}\times{10}$ $k$-mesh, is used to study the bulk electronic properties. Proportionate $k$-mesh is considered for slab calculation. The largest vector in the plane wave expansion is obtained by setting R$K_{max}$ = 7.0. For the fluorinated compound BaBiO$_2$F, the GGA optimized lattice parameter of 4.60 {\AA}, which is 0.25 {\AA} more than that of the pristine BaBiO$_3$, is used for the calculations.

The slabs, considered for the calculations, are of 15 units thick and grown along [001]. A vacuum of 15 {\AA} thick separates the consecutive slabs. Both top and bottom surfaces are taken to be Ba terminated (i.e. either BaF or BaO) as earlier studies have established the formation of TIDC with the Ba termination\cite{Clau}. The schematic representative of the slabs are shown in Fig.~\ref{fig:1}(b), (c). The surface band structure is calculated along the high-symmetry path of the surface BZ which is basically a projection of the bulk BZ in the $k_x$-$k_y$ plane (see Fig.~\ref{fig:1}(e)).  

\begin{center}
\begin{figure*}
\includegraphics[angle=-0.0,origin=c,height=10.0cm,width=15.0cm]{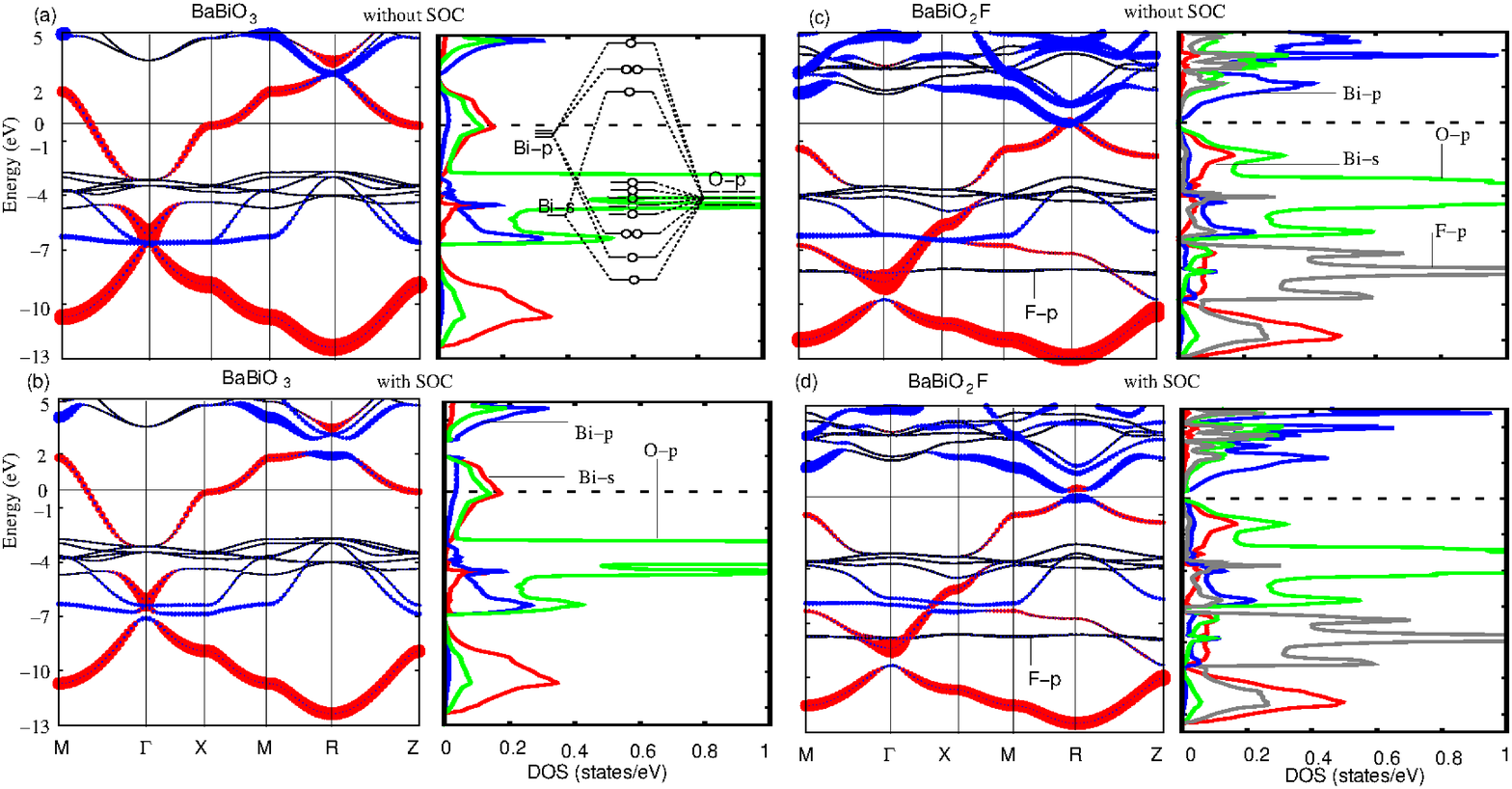}
\caption{The band structure of BaBiO$_3$ (a and b) and BaBiO$_2$F (c and d) without and with spin-orbit coupling. The Fermi level (E$_F$) is set to zero. The red and blue circles represent proportionate contribution of Bi-s and Bi-p states respectively towards forming the bands. The partial DOS, plotted on the right of each band structure further illustrates the orbital composition of the bands.  The molecular orbital picture\cite{Nanda} responsible for the formation of these bands is shown in the inset. For the pure compound, while the s-p band inversion occurs without SOC, the gap appearing at $\sim$ 2 eV above E$_F$ is SOC driven. For the fluorinated compound, both band inversion and band gap opening are SOC driven. Here, F provides the extra electron to shift the band gap to E$_F$.}
\label{fig:2}
\end{figure*}
\end{center}

The band structure and hence the transport properties of the pervoskites ABX$_3$ with X being a halogen or oxygen can be determined through the valence electron count (VEC)\cite{Kash}, i.e. the number of valence electrons per formula unit. The molecular orbital picture, drawn based on the DFT obtained band structure (Fig.~\ref{fig:2}a), suggests that if the VEC is 20, the bulk compound is a semimetal or semiconductor\cite{Kash}. Depending on the strength of spin-orbit coupling (SOC) and thereby the s-p band inversion, these semimetals/semiconductors give rise to the formation of TIDC at the surface boundary\cite{Nanda}. If the VEC differs from 20, the system is a metal and ceases to exhibit the TI phase.


 The contrasting transport properties of ABX$_3$, i.e. semimetal/semiconductor leading to TI behavior for VEC = 20 and metallic otherwise, is solely governed by a single salient feature of the band structure. An universal band gap (zero or finite) exists at the TRIM R ($\pi/a$,  $\pi/a$, $\pi/a$)  between the downward dispersive lower lying B-s dominated  singlet state and upward dispersive B-p dominated triplet states which can be noticed from Fig.~\ref{fig:2}a. These dispersive bands emerge out of the antibonding bands formed by the Bi-$\{$s,p$\}$- O-p chemical bonding, as schematically demonstrated by the molecular-orbital picture (inset of Fig.~\ref{fig:2}a). With VEC equals 20, the singlet state is completely occupied to form the band gap at E$_F$\cite{Kash}. For BaBiO$_3$, the VEC is 19 and hence the the singlet state is partially occupied to produce the (zero) band gap approximately 2 eV above E$_F$ (see Fig.~\ref{fig:2}a). The spin-orbit coupling amplifies the band gap (see Fig.~\ref{fig:2}b) within which time reversal symmetry protected Dirac states appear\cite{Nanda}. Fluorine with one additional valence electron compared to oxygen acts as a perfect electron donor. For BaBiO$_2$F, the VEC is 20 and therefore, the band gap appears at E$_F$ (Fig.~\ref{fig:2}d). Furthermore, the F-p states dominate the valence band spectrum in the range -6 to -10 eV w.r.t. E$_F$.  We may note that the family of ABiO$_3$ exhibits multiple Dirac states\cite{Nanda}, one in the valence band spectrum (around 7 eV below E$_F$), formed by the bonding Bi-$\{$s,p$\}$- O-p states,  and other in the conduction band spectrum (around 2eV above E$_F$) formed by the antibonding states. With substitution of fluorine, the F-p states dominate the bonding states (see Fig.~\ref{fig:2}c) instead of the Bi-p states.  Therefore, the s-p band inversion  does not occur (see Fig.~\ref{fig:2}d) and hence the TIDC is not formed in the valence band. On the other hand, the antibonding spectrum is unperturbed by the F-p states and hence neither the s-p band inversion nor the SOC driven band gap is affected by the fluorination as can be observed from Fig.~\ref{fig:2}d.  Hence, it is expected that the TIDC will form in the antibonding spectrum.

\begin{figure}
\begin{center}
\includegraphics[angle=-0.0,origin=c,height=6cm,width=7cm]{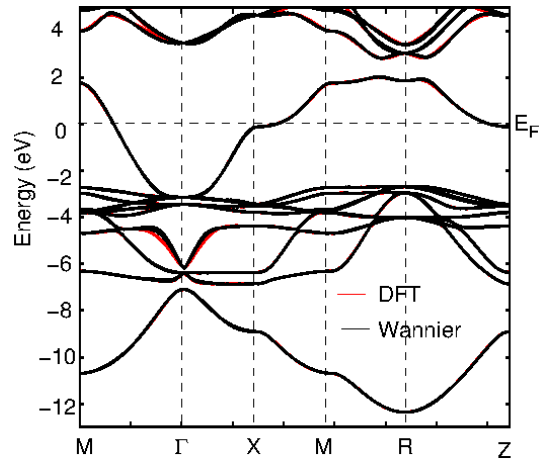}
\caption{Comparison of the band structure obtained from DFT (red) and from the TB model using MLWF basis(black).}
\label{fig:3}
\end{center}
\end{figure}

To verify the formation of TIDC for the pristine and fluorinated compounds, the surface states are calculated using Wannier formalism. First, the maximally localized Wannier functions (MLWF) as well as the strength of the hopping integral between these functions, are obtained from the bulk DFT calculations using Wannier90\cite{Mostofi}. Taking
MLWF as the elements of the basis, a tight-binding (TB) model
was employed on a slab structure to calculate the surface
Green's function through an iterative method\cite{Sancho,Lopez}as
implemented in WannierTools package\cite{arpes}. The appropriateness of the basis can be confirmed from the fact that the TB band structure has an excellent agreement with the full band structure obtained with DFT as can observed from Fig.~\ref{fig:3}. The imaginary part of the surface Green's function is the local density of state LDOS($k$, E).

\begin{center}
\begin{figure*}
\includegraphics[angle=-0.0,origin=c,height=10.0cm,width=15.0cm]{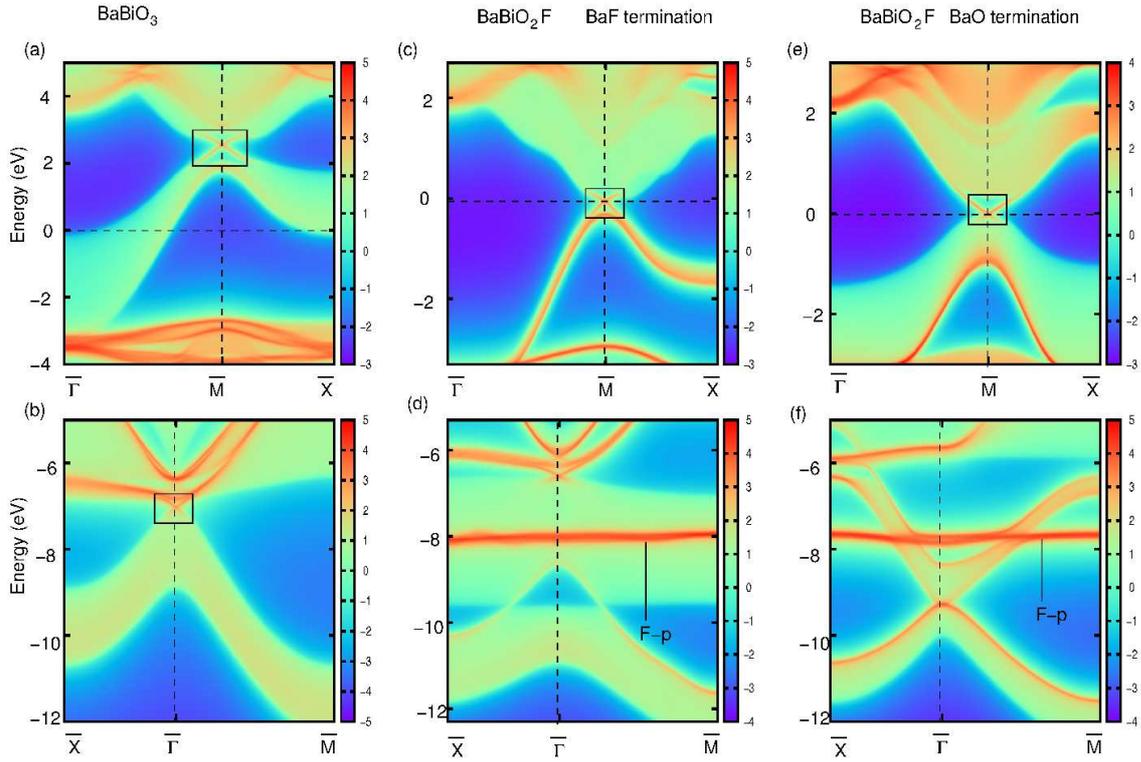}
\caption{The DFT+SOC simulated ARPES spectra (LDOS ($\kappa$, E)) of (001) surface for both BaBiO$_3$ and BaBiO$_2$F. E$_F$ is set to zero. The surface states are calculated using Wannier formalism for  BaBiO$_3$ (a, b),  BaBiO$_2$F with BaF termination (c, d) and  BaBiO$_2$F with BaO termination (e, f). The box highlights the formation of TIDC. With doping the E$_F$ shifts to the TIDC irrespective of the surface termination.}
\label{fig:4}
\end{figure*}
\end{center}

\begin{center}
\begin{figure}
\hspace{1.3cm}
\includegraphics[angle=-0.0,origin=c,height=10.5cm,width=14.0cm]{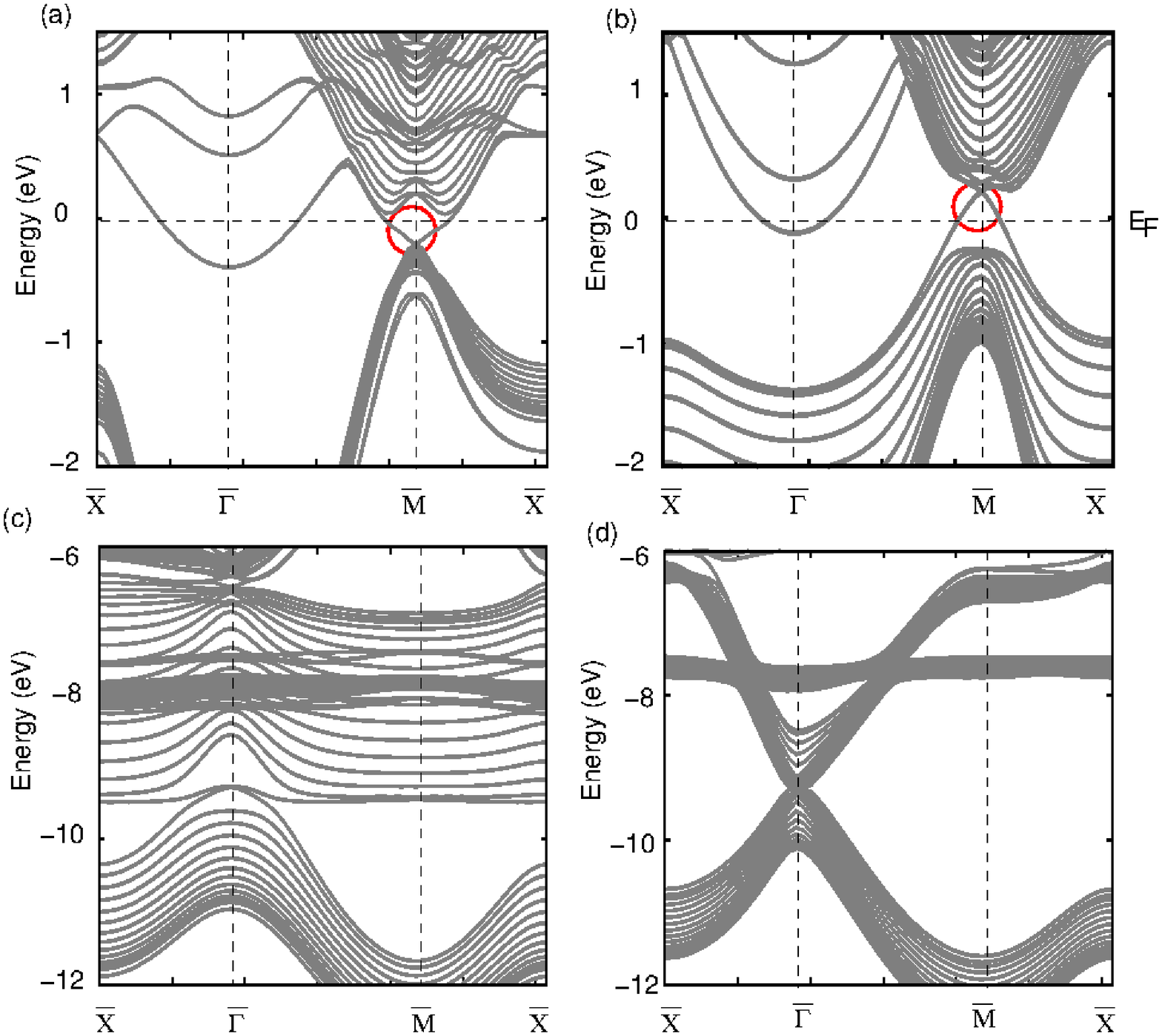}
\caption{The DFT + SOC obtained band structure of 15 units cell thick  BaBiO$_2$F slab with BaF termination (a, c) and  BaO termination (b, d).  The TIDC, indicated by red circle, is observed near E$_F$ for both the terminations.}
\label{fig:5}
\end{figure}
\end{center}

For the pristine and oxyfluoride [001] slabs (Fig.~\ref{fig:1}), the LDOS($k$, E) are plotted in Fig.~\ref{fig:4}. 
The colour gradient is a measure of LDOS. Deep red
and purple correspond to highest and lowest LDOS respectively. The
faded green color reflects the bulk band structure. As expected for the pristine system, the TIDC appears at $\sim$ 2 eV above E$_F$ and $\sim$ -7 eV below E$_F$ (see Fig.~\ref{fig:4}(a) and (b)). With fluorination,  the TIDC appears at E$_F$ irrespective of whether F replaces O from the BaO plane (Fig.~\ref{fig:4}c) or from the BiO$_2$ plane (Fig.~\ref{fig:4}e) with respect to the growth direction (see Fig.~\ref{fig:1}-b and c). However, unlike the pristine compound, the TIDC does not form in the valence band spectrum for the fluorinated systems as can be seen from the Fig.~\ref{fig:4}(d) and Fig.~\ref{fig:4}(f). It  is due to the absence of s-p band inversion as discussed in the bulk band structure.  For further confirmation, we calculated the full band structure of a 15 units cell thick BaBiO$_2$F slab using DFT, which is shown in Fig.~\ref{fig:5}, and observed that the TIDC, which is marked by the red circle, is formed very close to E$_F$.

\begin{center}
\begin{figure}
\hspace{1.2cm}
\includegraphics[angle=-0.0,origin=c,height=9.0cm,width=14.0cm]{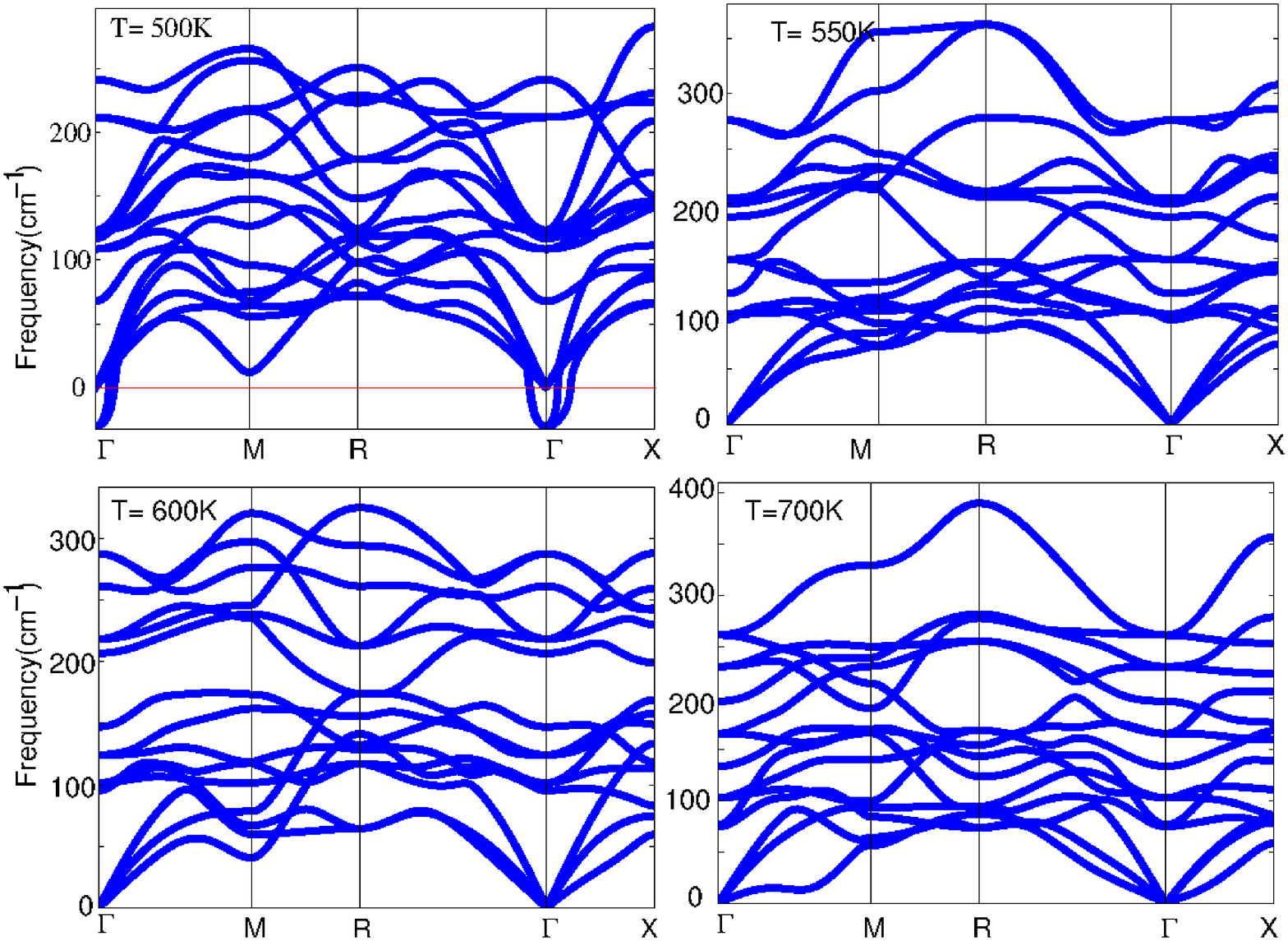}
\caption{Temperature dependent phonon spectra of BaBiO$_2$F. The thermodynamical stability of the high temperature cubic phase is ascertained as the negative frequencies are absent for T $>$ 500 K.}
\label{fig:6}
\end{figure}
\end{center}

 While the electronic structure of BaBiO$_2$F satisfies the necessary and sufficient criteria to form the TIDC near E$_F$,  the structural stability of such a system needs to be examined for its practical realization. The pristine BaBiO$_3$ exists in the cubic phase when the temperature is beyond $\sim$ 750 K.  \cite{Cox} Keeping this in mind,   we have  examined the  structural stability of the doped system in the temperature range 500 to 700K using the following two step process. In the first step, ab-initio molecular dynamics (AIMD) simulations were performed  on a $2\times2\times2$ supercell using VASP package\cite{Kresse,Joubert}. The PBE-GGA exchange-correlation functional was used and the plane wave cutoff energy was set to 500 eV. 
 The system was equilibriated at different temperatures (NVT) using velocity scaling method for ∼1.0 ps with a time step of 0.5 fs followed by ∼0.5 ps -long  production run  
 for collecting  the atomic displacement and force datasets. In the second step ALAMODE\cite{Tadano} package was used to get the harmonic and   interatomic force constants (IFCs). The phonon band dispersion were obtained by solving the dynamical matrix for the given wave vectors. The phonon dispersion curves for different temperatures are shown Fig.~\ref{fig:6}. The soft modes (negative frequencies) are observed when the temperature decreases below 500 K. Therefore, the cubic phase of these compounds exhibit dynamical stability for temperature greater than 500 K. By lowering the temperature, the soft modes introduce structural distortion. 
 
 The stability of the oxyfluoride is further quantified by calculating the formation and cohesive energies. They are estimated using the following expressions.
 \\
 
\begin{eqnarray}
E_{Form} &=& E_{BaBiO_{3-x}F_{x}} - E_{BaBiO_{3}}  -  x\frac{1}{2} E_{F_{2}} + x\frac{1}{2} E_{O_{2}},  \\
E_{Coh} &=&   E_{BaBiO_{3-x}F_{x}} - E_{Ba} - E_{Bi} - (3-x)\frac{1}{2} E_{O_{2}} -  x\frac{1}{2} E_{F_{2}}, 
\end{eqnarray}

Where $E_{BaBiO_{3-x}F_{x}}$, $E_{BaBiO_{3}}$ are total energy in their ground state. $E_{F_{2}}$ and $E_{O_{2}}$ are estimated by keeping respective molecules in a very large cubic box (size = 15 \AA ). While the $E_{Form}$ is found to be $-$ 0.69 eV, the  $E_{Coh}$ is found to be -2.45 eV/atom. The negative values indicates that the oxyfluoride can be synthesized independently or via substitution of Fluorine in BaBiO$_3$.

\begin{center}
\begin{figure}
\hspace{0.3cm}
\includegraphics[angle=-0.0,origin=c,height=9.0cm,width=15.0cm]{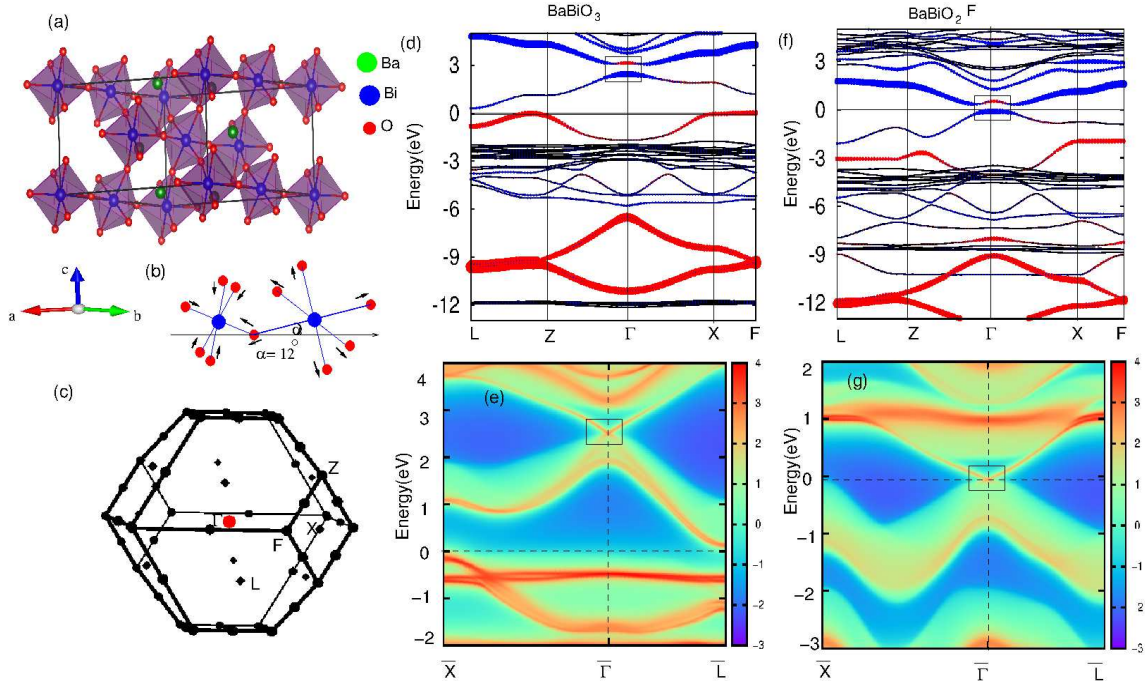}
\caption{(a) The monoclinic 
(C2/m) crystal structure of BaBiO$_3$. (b) Schematic illustration of  breathing and stretching mode of distortions, and tilting of octahedra observed in the monoclinic structure. (c) The Brillouin zone for the monoclinic lattice. (d) The bulk band structure and (e) the simulated (001) surface ARPES spectra for BaBiO$_3$. (f) and (g) are same as (d) and (e) but for BaBiO$_2$F. The proportionate contribution of Bi-s and Bi-p states in each band are represented by red and blue respectively. The s-p band inversion is highlighted by circles and the formation of TI states are highlighted by rectangles. The Fermi level is set to zero.}
\label{fig:7}
\end{figure}
\end{center}

Experimentally a series of structural transitions are observed in BaBiO$_3$.\cite{Howard}  The high temperature cubic phase yields to rhombohedral phase at 750K and at 405K the monoclinic structure (C2/m) forms the ground state . When the temperature goes below 104K, the compound stabilizes  with the primitive monoclinic structure with (P2$_{1}$/c).

 The room temperature monoclinic phase (see Fig.~\ref{fig:7}(a)) is obtained from the simultaneous occurring of the breathing and tilting distortions of the octahedra. 
 Interestingly, the earlier reported band structure calculations suggests that the TI Dirac state is protected against such distortions\cite{Clau}. The present work reconfirms the protection of the TI states as the bulk band structure for the monoclinic structure demonstrates s-p band inversion (see Fig.~\ref{fig:7}(d)) and the surface ARPES spectra demonstrates the formation of TI Dirac states (Fig.~\ref{fig:7}(e)).
 Anticipating a similar structural transition as in the pure compound, it is prudent to examine robustness of the TI state in the fluorinated system in the monoclinic (C2/m) phase.

 The bulk band structure of monoclinic BaBiO$_2$F is shown in Fig.~\ref{fig:7}(f). Like the pure compound (see Fig.~\ref{fig:7}(d)), the s-p band inversion is observed at $\Gamma$, but now at the Fermi level. As mentioned earlier, it is due to the fact that the participatory bands are shifted below by approximately 2.5 eV due to extra electron through F substitution. The ARPES spectra in Fig.~\ref{fig:7}(g) shows that the s-p band inversion leads to the formation of surface TI Dirac state in the monoclinic structure.

To summarize, our electronic structure calculations show that fluorine acts as a perfect electron donor to shift the Fermi level of the oxide perovskite BaBiO$_3$ without  affecting its symmetry protected surface Dirac states. For 33$\%$ fluorine substitution, the Dirac states lie exactly on the Fermi level to make the doped perovskite as a topological insulator. The thermodynamical stability of the fluorinated system beyond 500 K, as ascertained from phonon studies, suggests that the experimental realization of BaBiO$_3$ as topological insulator is possible through  doping. Universal nature of the band structure of cubic oxide perovskites implies that the finding of this work can be easily extended to other members of ABiO$_3$ where A is a divalent group-II element. Furthermore, we reveal that the TI state remain invariant with respect to the structural distortion that leads to the low temperature phase monoclinic structure.

\section{\label{sec:intro}Acknowledgement} The authors acknowledge HPCE, IIT Madras for providing the computational resources. This work is supported by DST, India through grant no. EMR/2016/003791. B. K. acknowledges K. Murali for computational help.

\section{References}
\bibliography{iop.bib}

\begin{thebibliography}{10}
\expandafter\ifx\csname url\endcsname\relax
  \def\url#1{{\tt #1}}\fi
\expandafter\ifx\csname urlprefix\endcsname\relax\def\urlprefix{URL }\fi
\providecommand{\eprint}[2][]{\url{#2}}

\bibitem{Hsieh}
Hsieh D, Xia Y, Qian D, Wray L, Meier F, Dil J~H, Osterwalder J, Patthey L,
  Fedorov A~V, Lin H, Bansil A, Grauer D, Hor Y~S, Cava R~J and Hasan M~Z 2009
  {\em Phys. Rev. Lett.\/} {\bf 103}(14) 146401
  \urlprefix\url{https://link.aps.org/doi/10.1103/PhysRevLett.103.146401}

\bibitem{Liang}
Fu L, Kane C~L and Mele E~J 2007 {\em Phys. Rev. Lett.\/} {\bf 98}(10) 106803
  \urlprefix\url{https://link.aps.org/doi/10.1103/PhysRevLett.98.106803}

\bibitem{Hasan}
Hasan M~Z and Kane C~L 2010 {\em Rev. Mod. Phys.\/} {\bf 82}(4) 3045--3067
  \urlprefix\url{https://link.aps.org/doi/10.1103/RevModPhys.82.3045}

\bibitem{Wray}
Lin H, Markiewicz R~S, Wray L~A, Fu L, Hasan M~Z and Bansil A 2010 {\em Phys.
  Rev. Lett.\/} {\bf 105}(3) 036404
  \urlprefix\url{https://link.aps.org/doi/10.1103/PhysRevLett.105.036404}

\bibitem{Pal}
Xia Yand~Qian D, Hsieh D, Wray L, Pal A, Lin H, Bansil A, Grauer D, Hor Y~S,
  Cava R~J and Hasan M~Z 2012 {\em Nat. Phys.\/} {\bf 5}(6) 398--402
  \urlprefix\url{http://dx.doi.org/10.1038/nphys1274}

\bibitem{Sato}
Sato T, Segawa K, Guo H, Sugawara K, Souma S, Takahashi T and Ando Y 2010 {\em
  Phys. Rev. Lett.\/} {\bf 105}(13) 136802
  \urlprefix\url{https://link.aps.org/doi/10.1103/PhysRevLett.105.136802}

\bibitem{Kubler}
Chadov S, Qi X, Kubler J, Fecher G~H, Felser C and Zhang S~C 2010 {\em Nat.
  Mater\/} {\bf 9}(7) 541 \urlprefix\url{http://dx.doi.org/10.1038/nmat2770}

\bibitem{Moore}
Moore J~E and Balents L 2007 {\em Phys. Rev. B.\/} {\bf 75}(12) 121306
  \urlprefix\url{https://link.aps.org/doi/10.1103/PhysRevB.75.121306}

\bibitem{Hughes}
Qi X~L, Hughes T~L and Zhang S~C 2008 {\em Phys. Rev. B.\/} {\bf 78}(19) 195424
  \urlprefix\url{https://link.aps.org/doi/10.1103/PhysRevB.78.195424}

\bibitem{Fu}
Fu L and Kane C~L 2007 {\em Phys. Rev. B.\/} {\bf 76}(4) 045302
  \urlprefix\url{https://link.aps.org/doi/10.1103/PhysRevB.76.045302}

\bibitem{Shuichi}
Murakami S, Nagaosa N and Zhang S~C 2004 {\em Phys. Rev. Lett.\/} {\bf 93}(15)
  156804 \urlprefix\url{https://link.aps.org/doi/10.1103/PhysRevLett.93.156804}

\bibitem{LiangFu}
Fu L and Kane C~L 2008 {\em Phys. Rev. Lett.\/} {\bf 100}(9) 096407
  \urlprefix\url{https://link.aps.org/doi/10.1103/PhysRevLett.100.096407}

\bibitem{David}
Shi H, Parker D, Du M~H and Singh D~J 2015 {\em Phys. Rev. Applied.\/} {\bf
  3}(1) 014004
  \urlprefix\url{https://link.aps.org/doi/10.1103/PhysRevApplied.3.014004}

\bibitem{Shou}
Yan B and Zhang S~C 2012 {\em Rep. Prog. Phys.\/} {\bf 75} 096501 ISSN
  0034-4885 (\textit{Preprint} \eprint{arXiv:1610.08983v2})
  \urlprefix\url{http://stacks.iop.org/0034-4885/75/i=9/a=096501?key=crossref.97671bd7fc614d8b61c261a9d446873e}

\bibitem{Cheng}
Zhang H, Liu C~X, Qi X~L, Dai X, Fang Z and Zhang S~C 2009 {\em Nat. Phys.\/}
  {\bf 5}(6) 438--442 \urlprefix\url{http://dx.doi.org/10.1038/nphys1270}

\bibitem{Park}
Park K, Heremans J~J, Scarola V~W and Minic D 2010 {\em Phys. Rev. Lett.\/}
  {\bf 105}(18) 186801
  \urlprefix\url{http://link.aps.org/doi/10.1103/PhysRevLett.105.186801}

\bibitem{Jun}
Xiao D, Yao Y, Feng W, Wen J, Zhu W, Chen X~Q, Stocks G~M and Zhang Z 2010 {\em
  Phys. Rev. Lett.\/} {\bf 105}(9) 096404
  \urlprefix\url{https://link.aps.org/doi/10.1103/PhysRevLett.105.096404}

\bibitem{Clau}
Yan B, Jansen M and Felser C 2013 {\em Nat. Phys.\/} {\bf 9}(11) 709--711
  \urlprefix\url{http://dx.doi.org/10.1038/nphys2762}

\bibitem{Binghai}
Yan B and Zhang S~C 2012 {\em Rep. Prog. Phys\/} {\bf 75} 096501

\bibitem{Jin}
Jin H, Rhim S~H, Im J, Freeman A~J and Fermions M 2013 {\em Sci. Rep.\/} {\bf
  3} 01651

\bibitem{Pi}
Pi S~T, Wang H, Kim J, Wu R, Wang Y~K and Lu C~K 2017 {\em J. Phys. Chem.
  Lett.\/} {\bf 8}(2) 332--339
  \urlprefix\url{http://dx.doi.org/10.1021/acs.jpclett.6b02860}

\bibitem{Thomale}
Li G, Yan B, Thomale R and Hanke W 2015 {\em Sci. Rep.\/} {\bf 5} 10435
  \urlprefix\url{http://dx.doi.org/10.1038/srep10435}

\bibitem{Nanda}
Khamari B, Kashikar R and Nanda B~R~K 2018 {\em Phys. Rev. B.\/} {\bf 97}(4)
  045149 \urlprefix\url{https://link.aps.org/doi/10.1103/PhysRevB.97.045149}

\bibitem{Cava}
Cava R~J, Batlogg B, Krajewski J~J, Farrow R, {Rupp Jr} L~W, White A~E, Short
  K, Peck W~F and Kometani T 1988 {\em Nature\/} {\bf 332} 814
  \urlprefix\url{http://dx.doi.org/10.1038/332814a0 http://10.0.4.14/332814a0}

\bibitem{Kanecl}
Fu L and Kane C~L 2008 {\em Phys. Rev. Lett.\/} {\bf 100}(9) 096407
  \urlprefix\url{https://link.aps.org/doi/10.1103/PhysRevLett.100.096407}

\bibitem{Hwan}
Song J~H, Jin H and Freeman A~J 2010 {\em Phys. Rev. Lett.\/} {\bf 105}(9)
  096403
  \urlprefix\url{https://link.aps.org/doi/10.1103/PhysRevLett.105.096403}

\bibitem{Linder}
Linder J, Tanaka Y, Yokoyama T, Sudb\o{} A and Nagaosa N 2010 {\em Phys. Rev.
  Lett.\/} {\bf 104}(6) 067001
  \urlprefix\url{https://link.aps.org/doi/10.1103/PhysRevLett.104.067001}

\bibitem{Qiran}
Li L~H, Cai Q, Shih C~j and Santos E~J~G 2018 {\em Nat. Commun.\/} {\bf 9} 1271
  \urlprefix\url{http://dx.doi.org/10.1038/s41467-018-03592-3}

\bibitem{Usui}
Usui H, Suzuki K and Kuroki K 2012 {\em Phys. Rev. B.\/} {\bf 86}(22) 220501
  \urlprefix\url{https://link.aps.org/doi/10.1103/PhysRevB.86.220501}

\bibitem{Lee}
Lee J, Stone M~B, Huq A, Yildirim T, Ehlers G, Mizuguchi Y, Miura O, Takano Y,
  Deguchi K, Demura S and Lee S~H 2013 {\em Phys. Rev. B.\/} {\bf 87}(20)
  205134 \urlprefix\url{https://link.aps.org/doi/10.1103/PhysRevB.87.205134}

\bibitem{Takumi}
Kamihara Y, Watanabe T, Hirano M and Hosono H 2008 {\em Journal of the American
  Chemical Society\/} {\bf 130} 3296--3297
  \urlprefix\url{http://dx.doi.org/10.1021/ja800073m}

\bibitem{Clemens}
Clemens O, Berry F~J, Wright A~J, Knight K~S, Perez-Mato J~M, Igartua J~M and
  Slater P~R 2013 {\em Journal of Solid State Chemistry\/} {\bf 206} 158--169
  ISSN 00224596 \urlprefix\url{http://dx.doi.org/10.1016/j.jssc.2013.08.013}

\bibitem{Heap}
Heap R, Slater P~R, Berry F~J, Helgason O and Wright A~J 2007 {\em Solid State
  Communications\/} {\bf 141} 467 -- 470 ISSN 0038-1098
  \urlprefix\url{http://www.sciencedirect.com/science/article/pii/S003810980601043X}

\bibitem{Frank}
Berry F~J, Coomer F~C, Hancock C, Ãrn Helgason, Moore E~A, Slater P~R, Wright
  A~J and Thomas M~F 2011 {\em Journal of Solid State Chemistry\/} {\bf 184}
  1361 -- 1366 ISSN 0022-4596
  \urlprefix\url{http://www.sciencedirect.com/science/article/pii/S0022459611001794}

\bibitem{Thompson}
Thompson C~M, Blakely C~K, Flacau R, Greedan J~E and Poltavets V~V 2014 {\em
  Journal of Solid State Chemistry\/} {\bf 219} ISSN 0022-4596
  \urlprefix\url{http://dx.doi.org/10.1016/j.jssc.2014.07.019}

\bibitem{Shim}
Berry F~J, Heap R, Helgason Ã, Moore E~A, Shim S, Slater P~R and Thomas M~F
  2008 {\em Journal of Physics: Condensed Matter\/} {\bf 20} 215207
  \urlprefix\url{http://stacks.iop.org/0953-8984/20/i=21/a=215207}

\bibitem{Kennedy}
Kennedy B~J, Howard C~J and Knight K~S 2006 {\em Acta Crystallogr. B.\/}
  537--546

\bibitem{Sle}
{Cox} D~E and {Sleight} A~W 1976 {\em Solid State Communications\/} {\bf 19}
  969--973 \urlprefix\url{http://adsabs.harvard.edu/abs/1976SSCom..19..969C}

\bibitem{Cox}
Cox D~E and Sleight A~W 1979 {\em Acta Crystallographica Section B\/} {\bf 35}
  1--10 \urlprefix\url{https://doi.org/10.1107/S0567740879002417}

\bibitem{Hamann}
Hamann D~R 1979 {\em Phys. Rev. Lett.\/} {\bf 42}(10) 662--665
  \urlprefix\url{https://link.aps.org/doi/10.1103/PhysRevLett.42.662}

\bibitem{Aug}
Blaha P, Schwartz K, Madsen G, Kvasnicka D and Luitz J 2001 {\em WIEN2k An
  Augmanted Plane Wave+Local Orbitals Program for Calculating Crystal
  Properties\/} (Karlheinz Schwartz, Tech. Universitt Wien, Austria)

\bibitem{Perdew}
Perdew J~P, Burke K and Ernzerhof M 1996 {\em Phys. Rev. Lett.\/} {\bf 77}(18)
  3865--3868
  \urlprefix\url{https://link.aps.org/doi/10.1103/PhysRevLett.77.3865}

\bibitem{Tran}
Tran F and Blaha P 2009 {\em Phys. Rev. Lett.\/} {\bf 102}(22) 226401
  \urlprefix\url{https://link.aps.org/doi/10.1103/PhysRevLett.102.226401}

\bibitem{Kash}
Kashikar R, Khamari B and Nanda B~R~K 2018 {\em Phys. Rev. Materials\/} {\bf
  2}(12) 124204
  \urlprefix\url{https://link.aps.org/doi/10.1103/PhysRevMaterials.2.124204}

\bibitem{Mostofi}
Mostofi A~A, Yates J~R, Lee Y~S, Souza I, Vanderbilt D and Marzari N 2008 {\em
  Computer Physics Communications\/} {\bf 178} 685--699 ISSN 0010-4655
  \urlprefix\url{http://www.sciencedirect.com/science/article/pii/S0010465507004936}

\bibitem{Sancho}
Sancho M~P~L, Sancho J~M~L and Rubio J 1984 {\em J. Phys. F: Met. Phys\/} {\bf
  14} 1205 \urlprefix\url{http://stacks.iop.org/0305-4608/14/i=5/a=016}

\bibitem{Lopez}
Sancho M~P~L, Sancho J~M~L, Sancho J~M~L and Rubio J 1985 {\em J. Phys. F: Met.
  Phys\/} {\bf 15} 851
  \urlprefix\url{http://stacks.iop.org/0305-4608/15/i=4/a=009}

\bibitem{arpes}
Wu Q, Zhang S, Song H~F, Troyer M and Soluyanov A~A 2018 {\em Comput. Phys.
  Commun.\/} {\bf 224} 405 -- 416 ISSN 0010-4655
  \urlprefix\url{http://www.sciencedirect.com/science/article/pii/S0010465517303442}

\bibitem{Kresse}
Kresse G and Furthm\"uller J 1996 {\em Phys. Rev. B\/} {\bf 54}(16)
  11169--11186
  \urlprefix\url{https://link.aps.org/doi/10.1103/PhysRevB.54.11169}

\bibitem{Joubert}
Kresse G and Joubert D 1999 {\em Phys. Rev. B\/} {\bf 59}(3) 1758--1775
  \urlprefix\url{https://link.aps.org/doi/10.1103/PhysRevB.59.1758}

\bibitem{Tadano}
Tadano T, Gohda Y and Tsuneyuki S 2014 {\em Journal of Physics: Condensed
  Matter\/} {\bf 26} 225402
  \urlprefix\url{http://stacks.iop.org/0953-8984/26/i=22/a=225402}

\bibitem{Howard}
Howard C~J, Kennedy B~J and Woodward P~M 2003 {\em Acta Crystallographica
  Section B\/} {\bf 59} 463--471
  \urlprefix\url{https://doi.org/10.1107/S0108768103010073}

\end{thebibliography}
\providecommand{\newblock}{}

\end{document}